\documentstyle{mn}

\input{epsf}

\newcommand{\be}{\begin{equation}}
\newcommand{\ee}{\end{equation}}
\newcommand{\bea}{\begin{eqnarray}}
\newcommand{\eea}{\end{eqnarray}}

\def\p{\partial}
\def\a{\alpha}

\def\r{\rho}
\def\g{\gamma}
\def\o{\omega}
\def\th{\theta}

\def\s{\sigma}
\def\ra{\rightarrow}
\def\Ra{\Rightarrow}

\def\M{{\cal M}}
\def\R{{\cal R}}

\font\cmss=cmss10 \font\cmsss=cmss10 at 7pt
\def\IZ{\relax\ifmmode\mathchoice
{\hbox{\cmss Z\kern-.4em Z}}{\hbox{\cmss Z\kern-.4em Z}}
{\lower.9pt\hbox{\cmsss Z\kern-.4em Z}}
{\lower1.2pt\hbox{\cmsss Z\kern-.4em Z}}\else{\cmss Z\kern-.4em Z}\fi}

\begin{document}

\title{Analytic solutions for spherical gravitational gas accretion
on to a solid body}

\author[Jos\'e Gaite]{Jos\'e Gaite\\
{\small\em Instituto de Matem{\'a}ticas y F{\'\i}sica
Fundamental,}
{\small\em CSIC, Serrano 123, 28006 Madrid, Spain}
}
\maketitle


\begin{abstract}
The process of gravitational accretion of initially homogeneous
gas onto a solid ball is studied with methods of fluid dynamics.
The fluid partial differential equations for polytropic flow can
be solved {\em exactly} in an early stage, but this solution soon
becomes discontinuous and gives rise to a shock wave. Afterwards,
there is a crossover between two intermediate asymptotic
self-similar regimes, where the shock wave propagates outwards
according to two similarity laws, initially accelerating, then
decelerating (and eventually vanishing).  Lastly, we study the final
static state. Our purpose is to attain a global picture of the
process.
\end{abstract}


\begin{keywords}
accretion -- hydrodynamics -- shock waves
\end{keywords}

\section{Introduction}

The problem of spherical infall of gas onto a solid object
has application in several astrophysical situations.
For example, it can represent the infall of gas onto a neutron star
from the surrounding cloud. Alternatively, it can be a simplified model
for the formation of planetary atmospheres as a result of gas
accretion onto the solid (rock) surface of a previously formed
spherical core.

Aspects of this fluid dynamical problem have been treated before
in the literature. A simplified formulation with constant gravity
was considered by Bisnovatyi-Kogan, Zel'dovich \& Nadezhin (1972)
and a self-similar solution with a shock wave found. Self-similar
solutions with the (variable) gravity due to a central point-like
mass, applicable to either ordinary or neutron stars, were studied
by Cheng (1977). Similarity methods favour power-law
distributions. This property was amply used in the study of
supernova explosion and later fallback by Chevalier (1989). An
extensive study of self-similar spherical accretion with an
initial power-law radial distribution of gas was provided by
Kazhdan \& Murzina (1994). They concluded that a
realistic solution should be an interpolation between a
self-similar solution with zero mass flux near the origin
(of which they derived general features) and a
solution with constant gravity and the correct boundary conditions
at the solid surface.

Chevalier (1989) and Kazhdan \& Murzina (1994) assume that the
incident gas flow is cold ($T=0$) and, respectively, that its
accretion rate decays with time with some exponent or that its
density decays with radius with exponent $\o <3$ (both conditions
are related). A cold inflow is not a good approximation for large
distances where the gas thermal energy is non-negligible with
respect to its gravitational energy. On the other hand, the
condition $\o <3$ implies that the mass of gas diverges at large
radius, making the neglect of its self gravity questionable.

One may try to amend these problems by endowing the incident gas
with pressure. For polytropic flow, similarity demands that the
initial pressure be also a power law with radius and, furthermore,
determines the exponents (in terms of the polytropic exponent).
This is the case studied by Cheng (1977) and it may seem somewhat
unnatural. Besides, the exponent of the density distribution also
leads to diverging mass with radius and, hence, to the
self-gravity problem.

In this paper, we relax the demand of having similarity at the
ouset and consider instead simple {\em uniform} initial conditions
for the gas. In other words, the initial configuration is a solid
body with a spherically symmetric mass distribution (a ball)
placed in a homogeneous gas with pressure, and the problem is to
analyze the subsequent evolution of this gas under the body's
gravity. The evolution will consist of infall of gas with a
progressive modification of the gas distribution around the body,
while the gas stays homogeneous far from it. This is essentially
the type of accretion treated by Bondi (1952), although the
artificial imposition of stationary flow allowed him to dispense
with differential equations. At any rate, a homogeneous
distribution of gas is just a particular case of power law ($\o =
0$) and, therefore, comparison with the results of Chevalier
(1989) and Kazhdan \& Murzina (1994) will be possible.

These simple initial conditions are suitable for an
analytic treatment, employing the full power of the methods of
nonlinear fluid dynamics. These allow us to attain a clear
picture of the initial non-trivial process near the surface. In
particular, the exact description of the formation of a shock wave
may illustrate its general features in accretion processes.
Even though the evolution is not self similar, similarity
arises in the course of it, and we shall see how and why.
Due attention is also paid
to the self-gravity question, and to
the decay towards a final static state.

This paper is organized as follows. In section \ref{magnitudes},
we introduce the relevant magnitudes in the problem and, hence,
various space and time scales, to reach a rough intuitive idea of
the physics involved and to define the limits of applicability of
the model. In section \ref{equations}, we introduce the fluid
equations to be used.  In section \ref{exact}, an exact solution
is found for the initial non-trivial dynamics, which occurs near
the ball's surface.  In the following section we obtain two
similarity solutions valid for larger $t$, the first still
confined within a short distance from the ball, while the second
is valid for large radius. Finally we consider the long-time
asymptotic static state and discuss the results.

A note on notation: we shall use frequently the asymptotic signs
$\sim$ and $\approx$; for example, $f(x) \sim g(x)$ or $f(x)
\approx g(x)$ (sometimes without making explicit the independent
variable $x$). The former means that $\lim f(x)/g(x)$ when $x$
goes to zero or infinity, as the case may be, is finite, while the
latter means, in addition, that the limit is one. On the other
hand, the sign $\sim$ may appear in some instances with the
related but looser meaning ``equal up to a numerical factor of the
order of unity".

\section{Relevant magnitudes and dimensional analysis}
\label{magnitudes}

The initial condition (the solid
ball in the homogeneous gas) is characterized by four parameters,
namely, the radius $R$ and mass $M$ of the ball, and the pressure
$P_0$ and density $\r_0$ of the gas (we assume that the gas is
perfect, inviscid, and non-heat-conducting or polytropic). In
addition, we have the constant of gravity $G$.  From these five
dimensional characteristic parameters, we can form two independent
non-dimensional numbers, namely, the ratio of densities $(4/3)\pi
R^3 (\r_0/M)$ and the ratio $R P_0/(\r_0 G M)$. The quantity
$P_0/\r_0 \sim c_0^2$ ($c_0$ being the sound speed) is
approximately the gas thermal energy per unit of mass. On the
other hand, $G M/R$ is the gas potential energy per unit of mass
on the ball. Their ratio measures the relative strength of gravity
in our problem.  To have any significant gas infall, we must
demand that the ball's gravity dominates over the gas thermal
energy, that is, $R c_0^2/(G M) \ll 1$.  The ratio of densities is
also very small. This is a necessary condition for neglecting the
self-gravity of the gas, as we will do, but it is not sufficient.
We shall discuss the sufficient condition after analysing the
typical length and time scales.

The basic length scale is $R$, of course. We can get a larger
length scale by dividing by the small number $R c_0^2/(G M)$:
\be
\R := {R\over R c_0^2/(G M)} =  {G M\over c_0^2}.
\ee
This length $\R$ marks the scale at which the gas thermal energy
is similar to its potential energy. It was introduced by Bondi
(1952) to define a non-dimensional radial variable. Alternatively,
we may divide $R$ by the cubic root of the other small number
$(4/3)\pi R^3 (\r_0/M)$:
\be
{R\over R[(4/3)\pi (\r_0/M)]^{1/3}} =
\left({3\over 4\pi}{M\over \r_0}\right)^{1/3},
\ee
in which we may suppress the numerical factor.%
\footnote{With an initial power-law density distribution
$\r = B\,r^{-\o}$, the analogous distance is $(M/B)^{1/(3-\o)}$.}
Obviously, this second length is the radius of a volume of gas
such that its mass is similar to $M$.

Analogously, we have a basic time scale, namely, $R^{3/2}/\sqrt{GM}$,
defined only in terms of the ball's parameters (and interpreted as
the typical time of Keplerian motion close to the ball's
surface). Dividing by $[R c_0^2/(G M)]^{3/2}$ we cancel the $R$
dependence and obtain $GM/c_0^{3}$, that is,
the time of sound propagation over
the distance $\R$ (note that the time of Keplerian motion at distance
$\R$ is $\R^{3/2}/\sqrt{GM} = GM/c_0^{3}$ as well). Finally, dividing
by $[R^3 (\r_0/M)]^{1/2}$, we cancel both the $R$ and $M$ dependences
and obtain $1/\sqrt{G\r_0}$, the typical time of gravitational
collapse of the gas (aside from the ball, which might act as a seed
for the collapse).

We will assume that the largest typical length (the radius of a mass
of gas $M$) is much larger than the intermediate scale $\R$;
in other words, we consider
the mass of gas in the volume of radius $\R$ negligible in
comparison with $M$.  Then, analogously, the typical time of gas
collapse is much larger than $GM/c_0^{3}$. This is the precise
condition for neglecting the self-gravity of the gas, understood
in an asymptotic sense, namely, $\R \ll (M/\r_0)^{1/3}$ or
$GM/c_0^{3} \ll 1/\sqrt{G\r_0}$.  In non-dimensional form,
$c_0^6/(G^3 M^2 \r_0) \gg 1$, which can be restated in terms of
the two previously defined non-dimensional numbers: besides that
both must be very small, the ratio
\be
{R^3 \r_0\over M}:\left({R c_0^2\over GM}\right)^3  =
\left({\R\over R}\right)^3 {R^3 \r_0\over M} \ll 1, \label{cond}
\ee
that is, the ratio of densities must be much smaller than the
other one.  Interestingly, under this condition, the parameters
$M$ and $G$ will not appear independently in the equations of
motion but only in the combination $G M$, so that one has only
{\em four} dimensional parameters. Correspondingly, only one
non-dimensional number is relevant, namely, the ratio $R/\R$.

As regards mass scales, the basic mass is $M$, of course. The mass
of gas enclosed in the sphere of radius $\R$, namely, $\M \approx
(4/3)\pi \R^3 \r_0$, can be interpreted as the total mass of
initially bound gas. We observe that $\M/M \sim \R^3 \r_0/M \ll
1$, according to condition (\ref{cond}). On the other hand, this
condition can also be written as $M_J \gg M$, introducing the gas
Jeans mass $M_J \sim c_0^3/(G^3 \r_0)^{1/2}$. This is, of course,
the natural mass scale for gas collapse due to self gravity.

It is convenient to remark that, since we still have after
neglecting the gas self gravity one non-dimensional number
$(R/\R)$, we can construct many length and time scales (large or
small), but they have no particular physical meaning. However,
multiplying the basic length $R$ by that number, we obtain $R^2
c_0^2/(G M) = c_0^2/g$, where $g$ is the gravity on the ball's
surface.  This small length scale and its associated time scale
$c_0/g$ will play a r\^ole in the sequel.

\section{Fluid equations}
\label{equations}

Under the assumption of spherical symmetry, we are led to
solving the partial differential equations of fluid dynamics in one
dimension (the radial distance). These equations are nonlinear and no
general method of solution is available. However, given the simplicity
of the initial and boundary conditions, many results can be obtained
by purely analytic means, as we shall see.

We consider an adiabatic evolution of the gas or, more generally,
a {\em polytropic} equation of state, $P \propto \rho^\g$ ($\g
\geq 1$). Then, we have the continuity equation, the Euler
equation and the thermodynamic equation:
\bea
{\p\r\over\p t} + {\p(\r v)\over\p x} + {2(\r v)\over x}=0, \label{cont}\\
{\p v\over\p t} + v {\p v\over\p x} = -g - {1 \over\r}
{\p P\over\p x},  \label{Eul}\\
{\p\over\p t}{P\over\r^\g} + v {\p \over\p x}{P\over\r^\g} = 0,
\label{thermo}
\eea
where $g = GM/x^2$ is the gravity acceleration (Chevalier 1989;
Kazhdan \& Murzina 1994). The initial conditions are: $\r(0,x) =
\r_0$, $P(0,x) = P_0$ and $v(0,x) = 0$ ($x \geq R$). The boundary
condition at the solid surface is $v(t,R) = 0$.

In principle, Eq.~(\ref{thermo}) has the trivial solution
$P/\r^\g = P_0/\r_0^\g$. Let us introduce the sound velocity $c$,
which satisfies
\be
c^2(\r)/c_0^2 = (\r/\r_0)^{\g-1}\quad (c_0^2 = \g P_0/\r_0).
\label{dens-c}
\ee
Hence, the mathematical problem boils down to solving
\bea
{\p\r\over\p t} + {\p(\r v)\over\p x} + {2(\r v)\over x}=0, \label{cont2}\\
{\p v\over\p t} + v {\p v\over\p x} = -g - {c^2(\r) \over\r}
{\p\r\over\p x}.  \label{Eul2}
\eea

The polytropic gas flow in one dimension is amenable to powerful
mathematical methods (Courant \& Friedrichs, 1948; Landau \&
Lifshitz 1987; Chorin \& Marsden 1993), although the presence of
gravitation complicates the problem. However, restricting our
interest to a small zone near the surface, we can take the
acceleration of gravity $g$ constant in Eq.~(\ref{Eul2})
(Bisnovatyi-Kogan, Zel'dovich \& Nadezhin 1972). This will allow
us to take advantage of those methods.

\section{Initial stages}
\label{exact}

Initially, the gas will start falling with acceleration $g$, that
is, with a negative velocity increasing as $v \approx -g\,t$.
However, it will be stopped at the ball's surface, where the
density and, therefore, the pressure must increase.  Consequently,
a wave transmitting the boundary condition $v(t,R) = 0$ will
propagate outwards. It is therefore crucial to determine the law
of propagation of waves in the present conditions. This is called,
in mathematical terms, the analysis of {\em characteristics}. Once
this is done, and as long as the dynamics consists of the
propagation of a {\em simple wave}, one can obtain an exact
solution. It is not possible here to describe in detail how this
solution is obtained and our focus will be the origin of the shock
wave. We refer the reader to Courant \& Friedrichs (1948), Landau
\& Lifshitz (1987) or Chorin \& Marsden (1993) for a general
treatment of the theory of one-dimensional gas flow.

For the moment, we confine ourselves to a spherical shell over the
surface of height much smaller than $R$,
where we can consider $g$ constant.
Hence, the problem
becomes that of the fall of gas on a flat surface (the ground)
and it is convenient to take this surface as the origin of $x$.
So we must use the one-dimensional form of the vector divergence
and, therefore, we must neglect the last term
on the left-hand side of the continuity equation
(\ref{cont2}), writing it as
\be
{\p\r\over\p t} + {\p(\r v)\over\p x} =0. \label{cont1}
\ee

The gas still unperturbed by the wave merely falls with velocity
$v=-g\,t$. In a reference system falling with it, it is at rest
and, therefore, the speed of sound is $c_0$. Hence, in the
original reference system the wave front is located at $x_f = c_0
t - g\, t^2/2$. Then, for $x > x_f$, the solution is trivial, and
we only need to find the solution for $x < x_f$. In the falling
frame, the acceleration of gravity vanishes and the set of two
equations (\ref{cont1}) and (\ref{Eul2}) reduces to a ``gas tube
problem" that can be solved by introduction of the Riemann
invariants (Courant \& Friedrichs, 1948; Landau \& Lifshitz 1987;
Chorin \& Marsden 1993). Let us denote $x',v',c'$ the coordinate
and variables in this frame. Initially, we have a {\em constant
state}, which is preserved for $x' \geq x'_f = c_0 t$. The
backward characteristics crossing the forward characteristic $x' =
c_0 t$ transmit a constant value of the Riemann invariant $J_-$,
so the solution is a forward simple wave. Given that
$$J_- = v'  - \frac{2c'}{\g-1} = - \frac{2c_0}{\g-1},$$
the sound velocity is related with the gas velocity by $c' = c_0 +
(\g-1)v'/2$. Furthermore, the fact that the solution is a forward
simple wave implies that the forward characteristics are straight
lines. Hence, the wave's propagation law $v'=F(x-c'\,t)$ is an
implicit equation for $v'$, assuming that we can determine the
function $F$. This is done using the boundary condition $v(t,0) =
0$. The implicit equation is algebraic and its solution is
straightforward. We then obtain:
\be v(x,t) =\left\{ {\begin{array}{l}
-gt, \quad x \geq c_0 t - {g\over 2} t^2\\

         -{c_0\over \g} - {\g-1\over 2\g} gt + \sqrt{\left({c_0\over
      \g} + {\g-1\over 2\g} gt\right)^2 - {2g\over\g}x},\\
           \hbox{\hspace{4cm}} x \leq
      c_0 t - {g\over 2} t^2. \end{array}} \right.
\label{vel}
\ee
For a simple wave, the density is a definite function of the
velocity. We can calculate it from Eq.\ (\ref{dens-c}) to be
\be \r(x,t) = \r_0 \left[1+{\g-1\over 2}\,{v(x,t)+gt\over
c_0}\right]^{{2\over\g-1}}.     \label{dens} \ee
These expressions for $\r$ and $v$ constitute the solution of
Eqs.~(\ref{Eul2}) and (\ref{cont1}) with the given boundary
conditions.

\begin{figure}
  \epsfxsize=8cm
\epsfbox{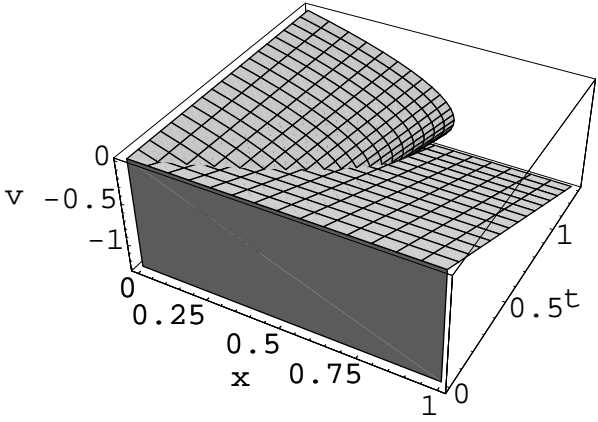}
\caption{Evolution of the velocity in the early stages in
non-dimensional units ($c_0 = g = 1$) for $\g=7/5$, displaying how
it becomes multivalued, hence giving rise to a shock wave.} 
\label{plotexact}
\end{figure}

The alert reader may have noted a peculiarity of the solution
found above: the wave front's coordinate $x_f$ begins increasing
but, for $t > c_0/g$, changes to decreasing and, eventually, returns
to the origin (free fall with initial velocity $c_0$).  On the
other hand, a simple analysis shows that a singularity occurs
before, namely, a shock wave.  One can calculate $(\p x/\p v)_t$,
and its null locus defines a line in the $xt$-plane, namely, $x =
[2c_0 + (\g-1)g t]^2/(8\g g)$. Initially, the corresponding $x$ is
larger than $x_f$, so it is unphysical, but, for
$t=2c_0/[(\g+1)g]$, it meets the wave front; that is, the falling
gas overtakes the gas just below it and a shock wave develops
henceforth. From that moment onwards, the solution
(\ref{vel},\ref{dens}) becomes multivalued and ceases to be valid
(see Fig.~\ref{plotexact}). Although it is possible, in principle,
to study the situation in which a shock wave propagates, the
analysis cannot be done in terms of a simple wave and it is far
more complicated.

Of course, the foregoing analysis is valid as long as
everything happens in a thin
shell, which implies that the typical length $c_0^2/g \ll R$, equivalent to
$\R \gg R$ (the already mentioned
condition of gravity dominated evolution).

\section{Similarity solutions}
\label{simil}

A one-dimensional gas flow problem may be formulated with
similarity variables [in which the fluid partial differential
equations become ordinary differential equations (ODEs)] only when
the initial and boundary conditions are sufficiently simple (Sedov
1982). Certainly, this is not our case, for we can construct
independent length and time scales (see section \ref{magnitudes}).
Nevertheless, it is commonly observed that nonlinear equations
that do not have similarity solutions at the outset develop them
in an {\em intermediate asymptotic} regime, that is, a regime
between two very different scales where both scales can be
neglected (Barenblatt 1996). We will actually find two independent
intermediate asymptotic regimes.

We shall study first the self similar solution that develops from
the exact solution found in section \ref{exact}, that is, with
constant gravity. The corresponding ODE have been formulated by
Bisnovatyi-Kogan, Zel'dovich \& Nadezhin (1972). However, we shall
derive a slightly different (though equivalent) system of
equations more amenable to analytic study.

Afterwards, we proceed to consider the variation of gravity with
radius, connecting with the similarity solutions of Chevalier
(1989) and Kazhdan \& Murzina (1994). These solutions are for an
initial density distribution given by $\r = B\, x^{-\omega}$ but a
constant density is just a particular case. The most interesting
solutions is the one with zero mass flux near the origin, named
``of type 3" (Kazhdan \& Murzina 1994), since it is expected to
match the self-similar solution with constant gravity.

\subsection{Constant gravity}

In this case, we have one less parameter, because the parameters
$R$ and $GM$ only appear in the combination $g=GM/R^2$. However,
we can still form the two {\em independent} non-dimensional
variables $gt/c_0$ and $gx/c_0^2$, so we will have to remove
another parameter to have a similarity solution. If we assume that
the initial gas temperature is very low, then $c_0 \rightarrow 0$
and the only possible non-dimensional variable is $\xi = x/(g
t^2)$ (or a function of it). (We measure the distance $x$ from the
ground.) This corresponds to the intermediate asymptotics $c_0^2/g
\ll x \ll R$; equivalently, in terms of the non-dimensional height
variable $\tilde{x} = gx/c_0^2$, it corresponds to $1 \ll
\tilde{x} \ll \R/R $ (recall that $\R \gg R$).

To take into account the presence of shock waves and the
consequent dissipation, we consider the full set of equations
(\ref{Eul}), (\ref{thermo}) and (\ref{cont1}). We can express them
in non-dimensional form by introducing new variables: \be v =
{x\over t} u, \quad \r = \r_0 r, \quad P = {x^2\over
t^2}\,\r_0\,p. \ee Some straightforward algebra then yields, \bea
\xi[u' + (u-2) {r'\over r}] + u = 0,\\
(u-2)\,\xi u' + u^2 - u = -\xi^{-1} -r^{-1}(2p + \xi p') ,\\
(u-2)\,\xi [\ln(p/r^\g)]' + 2(u-1) = 0.
\eea
Upon making the change of variables
\be
\left\{
      {\begin{array}{l}
         \xi = \tilde\xi -{1\over 2},\\
        u = {\tilde\xi\tilde u -1\over \xi}, \\
        p =  {\tilde\xi^2 \tilde p\over \xi^2} \end{array}} \right.
\label{change}
\ee
(corresponding to changing to the falling reference frame),
the  $\xi^{-1}$ term in the second equation vanishes and
the system of ordinary differential equations
simplifies to 
\bea
\tilde{\xi}[\tilde{u}' + (\tilde{u}-2) {r'\over r}] + \tilde{u} = 0,\\
(\tilde{u}-2)\,\tilde{\xi} \tilde{u}' + \tilde{u}^2 - \tilde{u} = -r^{-1}(2\tilde{p} + \tilde{\xi} \tilde{p}') ,\\
(\tilde{u}-2)\,\tilde{\xi} [\ln(\tilde{p}/r^\g)]' + 2(\tilde{u}-1)
= 0.
\eea

Bisnovatyi-Kogan, Zel'dovich \& Nadezhin (1972) do not make the
change to the falling reference frame, and restrict themselves to
solving the equations by a series expansion near the ground.
Instead, after the change of reference, we can follow the general
(and more powerful) methods exposed by Sedov (1982). In
particular, we note that in terms of the variable $\tau = \ln
\tilde\xi$, the previous equations constitute a system of
autonomous nonlinear ODE, namely,
\bea
{d\tilde{u}\over d\tau} = {\frac{\tilde{p}(\tau)\,\left[ 2 +
\g\,\tilde{u}(\tau)
      \right] - r(\tau)\,\tilde{u}(\tau)\left( 2 - 3\,\tilde{u}(\tau) +
     {{\tilde{u}(\tau)}^2} \right) } {-\g\,\tilde{p}(\tau) + r(\tau)\,{{\left[
     \tilde{u}(\tau) - 2  \right] }^2}}}, \label{ODE1}\\
{dr\over d\tau} =  r(\tau){\frac{-2\,\tilde{p}(\tau) +
       r(\tau)\,\left[ \tilde{u}(\tau) -2 \right] \,\tilde{u}(\tau)}{
     \left( -\g\,\tilde{p}(\tau)   +
       r(\tau)\,{{\left[  \tilde{u}(\tau) -2 \right] }^2} \right) \,
     \left[\tilde{u}(\tau)-2 \right] }}, \label{ODE2}\\
{d\tilde{p}\over d\tau} =
\tilde{p}(\tau){\frac{2\,\g\,\tilde{p}(\tau) -
        r(\tau)\,\left( 4 - \left( 6 + \g \right) \,\tilde{u}(\tau) +
           2\,{{\tilde{u}(\tau)}^2} \right)}{-\g\,\tilde{p}(\tau) +
      r(\tau)\,{{\left[ \tilde{u}(\tau)-2 \right] }^2}}}.
      \label{ODE3}
\eea
One can study this system of equations with the methods of
nonlinear ODE, that is, study its singular points, phase portrait,
etc. However, given the homogeneity properties of these equations
with respect to $\tilde p$ and $r$, it proves convenient to
introduce the variable $\tilde\th= \tilde p/r$. It is (in the
falling frame) the non-dimensional form of the temperature for the
perfect gas with pressure $P$ and density $\r$, that is, $T =
\mu\,(x^2/t^2)\,\th$ ($\mu$ being the molar mass and $T$ being
measured in energy units). Hence,
\be
{d\tilde\th\over d \tilde{u}} = \tilde\th\,
{\frac{2\,\tilde\th\,\left( 1 + \g\,(\tilde{u}-2) \right)  -
     \left( \tilde{u} - 2\right) \,
      \left( 4 - \left( 5 + \g \right) \,\tilde{u} +
        2\,{{\tilde{u}}^2} \right) }{\left( \tilde{u} - 2 \right) \,
     \left( \tilde\th\,\left( 2 + \g\,\tilde{u} \right)
           - \tilde{u}\,\left( 2 - 3\,\tilde{u} + {{\tilde{u}}^2}
          \right)  \right) }} \label{ODE}
\ee
The solution of this equation provides the relation between the
``velocity'' $\tilde u$ and the ``temperature'' $\tilde\th$,
$\tilde\th({\tilde u})$. Substituting it back into
Eq.~(\ref{ODE1}), we have an ODE for ${\tilde u}(\tau)$, which is
immediately solved by a quadrature. Analogously, one solves for
$r(\tau)$.

To solve the nonlinear ODE (\ref{ODE}), let us note that we are
actually interested in some particular initial conditions, derived
from the initial and boundary conditions of the original partial
differential equations (PDE). The initial conditions for the PDE
are given at $t=0$, that is, at $\tilde\xi = \xi = \infty$. They
are $\r(0,x) = \r_0$, $P(0,x) = P_0 = 0$ ($c_0 \ra 0$) and $v(0,x)
= 0$, implying that $r=1$ and $u=0$. The boundary condition for
the PDE is given at $x=0 \Rightarrow \xi=0$ and $\tilde\xi = 1/2$,
where $v=0$.
Since $v = gt(\tilde\xi \tilde u-1)$, it implies that $\tilde u
=2$.

\begin{figure}
  \epsfxsize=8cm
\epsfbox{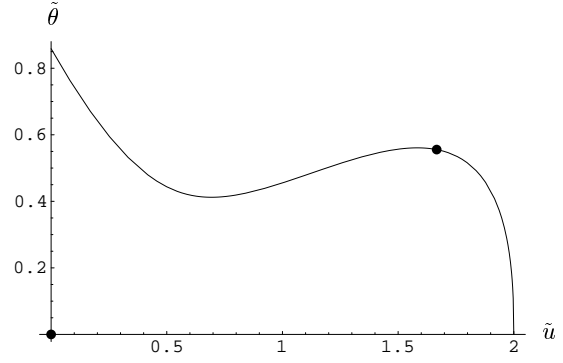} 
\caption{Relevant solution of the ODE (\ref{ODE}) for $\g = 7/5$,
displaying the origin and its image at the shock surface, the
point $(5/3,5/9)$.} 
\label{traject}
\end{figure}

Let us see if we have gathered enough information to determine a
unique solution of Eq.~(\ref{ODE}).  Given that $\tilde\th \ra 0$
as $\tilde\xi \rightarrow \infty$, this must be a fixed point.
There are three singular points of Eq.~(\ref{ODE}) with
$\tilde\th=0$, namely, with $\tilde u = 0,$ 1 or 2, respectively.
If we want to have a finite $v$ as $x \rightarrow \infty$, we must
take the point $(0,0)$.  So we know that the solution of
Eq.~(\ref{ODE}) that we seek must end at the origin. On the other
hand, the singular point $(2,0)$ corresponds to the boundary
conditions at $x=0$.  However, the solution that departs from this
point does not end at the origin (see Fig.~\ref{traject}).  This
does not mean that there is an inconsistency, for there can be
discontinuities in the solution. Indeed, in most self-similar
solutions the boundary conditions can only be fulfilled if there
is a point of discontinuity (Sedov 1982).

Therefore, recalling the exact solution of section \ref{exact}, we
observe that a shock wave arises and propagates upwards. Then we
must consider discontinuities associated to the presence of a
shock wave. The conditions for conservation of mass, momentum and
energy across a shock surface read (Courant \& Friedrichs, 1948;
Landau \& Lifshitz 1987; Chorin \& Marsden 1993)
\be \left\{
      {\begin{array}{l}
         \r_1 v_1 = \r_2 v_2,\\
        P_1 + \r_1 v_1^2 = P_2 + \r_2 v_2^2, \\
        {v_1^2\over 2} + w_1 = {v_2^2\over 2} + w_2, \end{array}} \right.
\label{shock}
\ee
where subindices refer to the values on each side of the shock and
$w$ is the enthalpy (or the appropriate thermodynamic function for
general $\g$); for a perfect gas, $w = \g P/[\r (\g-1)]$. These
equations hold in a coordinate system in which the shock surface
is at rest. On the other hand, the location of the shock must be
at fixed $\xi$, say $\xi_s$. Consequently, the shock wave velocity
is $$v_s = {dx_s \over dt} = \xi_s g {dt^2 \over dt} = 2\xi_s g t
=  2 {x_s \over t},$$ that is, $u_s = 2$. Then,
\be \left\{
      {\begin{array}{l}
         r_1 (u_1 -2) = r_2 (u_2 -2),\\
        p_1 + r_1 (u_1-2)^2 = p_2 + r_2 (u_2-2)^2, \\
        {(u_1-2)^2\over 2} + {\g \over \g-1}{p_1 \over r_1} =
       {(u_2-2)^2\over 2} + {\g \over \g-1}{p_2 \over r_2},\end{array}} \right.
\label{shock2}
\ee
valid in both the rest and the falling reference frames. Using the
falling frame and eliminating $r_2/r_1$ between the first and
second equations,
\be
\left\{
      {\begin{array}{l}
        {\tilde\th_1\over \tilde{u}_1-2} + \tilde{u}_1-2 =
        {\tilde\th_2\over \tilde{u}_2-2} + \tilde{u}_2-2, \\
        (\tilde{u}_1-2)^2 + {2\g \over \g-1}\tilde\th_1 =
       (\tilde{u}_2-2)^2 + {2\g \over \g-1}\tilde\th_2. \end{array}} \right.
\label{shock3} \ee

We have arrived at an involutive mapping of the plane $({\tilde
u},\tilde\th)$ as the relation between the variables at either
side of the shock discontinuity. Since we have the condition that
as $x \rightarrow \infty$ the solution goes to $(0,0)$ and (in the
falling frame) these values hold all the way down to the shock
surface, we just need the image of the origin under the mapping.
This yields the point
$$({\frac{4}{\g + 1}}, {\frac{8\,\left(\g - 1\right)} {{{\left(\g + 1
\right) }^2}}}).$$ Therefore, we need the solution of
Eq.~(\ref{ODE}) that goes through this point. We can see in an
example (Fig.~\ref{traject}) that the strand of this solution that
we need ends at the point (2,0), hence satisfying the
boundary condition at $x=0$.%
\footnote{Notice that this implies that the asymptotics $c_0
\rightarrow 0$ is of the {\em first kind} (the simple case)
according to the denomination of Barenblatt (1996).}

Now, substituting the function just computed back into Eq.~(\ref{ODE1}), we
obtain
\be d\tau = {\frac{\g\,{\tilde\th}(\tilde{u}) - {{\left( \tilde{u}
- 2
     \right) }^2}}{-\th(\tilde{u})\,\left( 2 + \g\,\tilde{u}
     \right) + \tilde{u}\,\left( 2 - 3\,\tilde{u} +
     {{\tilde{u}}^2} \right) } }\,d\tilde{u},
\ee
from which we can deduce the interval of $\tau$ between the points
$\left(4/(\g+1), 8\left(\g-1 \right)/{\left(\g+1 \right)
}^2\right)$ and (2,0) by integration. The result for $\g = 7/5$
(the adiabatic index of a perfect diatomic gas) is 0.0880104.
Hence, we obtain the quotient between the corresponding values of
$\tilde\xi$. This quotient gives the location of the shock
relative to the ground: recovering the original notation that
distinguishes both reference frames, ${\tilde\xi}_s/(1/2) = \exp
0.0880104 = 1.09200 \Rightarrow \xi_s = 0.0919995/2 = 0.0459997$,
so that the coordinate of the shock is $x_s = 0.0459997\, g t^2$.

\begin{figure}
  \epsfxsize=8cm
\epsfbox{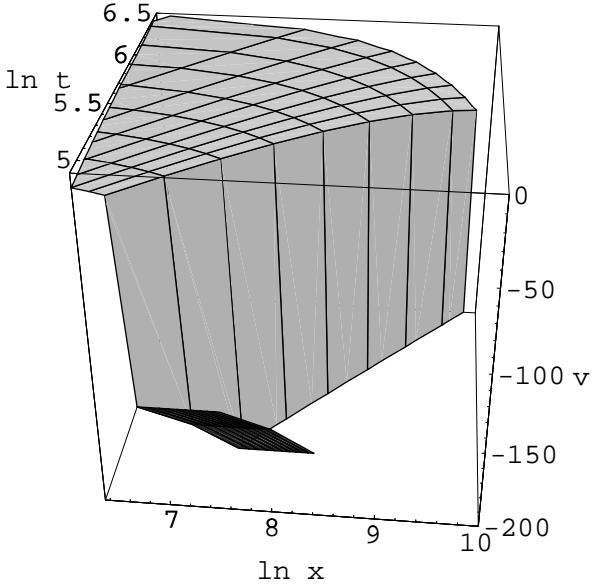} \caption{Self-similar velocity plot ($c_0 = g
= 1,$ $\g=7/5$) showing two sets of lines, for constant $t$ or
$\xi$, respectively (the border corresponds to the shock at
$\xi_s$) . The $v$ profile at constant $t$ is transported
homologously along the constant $\xi$ lines.}
\label{ssplot}
\end{figure}

It is easy to compute the post-shock $u$ in the rest frame: $u =
[(1.092/2)(5/3) - 1)]/0.046 = -1.957$. Hence, the post-shock
velocity is $v = (x_s/t)\,u = \xi_s u\, gt = -0.09 gt$. Comparing
with the pre-shock value $v = - gt$, we see a great reduction. The
corresponding kinetic energy is dissipated as heat. The similarity
properties of the velocity and its jump at the shock are shown in
Fig.~\ref{ssplot}.

The ratio of post-shock to pre-shock densities can be derived from
the first Eq.\ (\ref{shock2}) as $\r_2/\r_1 = (\g-1)/(\g+1) = 6$,
as corresponds to ``strong shocks" (Landau \& Lifshitz 1987). The
post-shock temperature is given by $T = \mu (x_s^2/t^2) \th =
\mu\left(8\left(\g-1 \right)/{\left(\g+1 \right)
}^2\right){\tilde\xi}_s^2 g^2 t^2 = 0.166\, \mu g^2 t^2 \Ra T/T_0
= 0.232\, (gt/c_0)^2$. Therefore, the post-shock pressure, $P =
\r\, T/\mu$, is not polytropic.

Other properties of the self-similar solution found are worth
noting, but require further analysis of the ODE system
(\ref{ODE1}, \ref{ODE2}, \ref{ODE3}). For example, at the ground,
$\tilde\th = \xi^2\th/{\tilde\xi}^2 = 0$, so that $\th = \lim_{\xi
\ra 0} \tilde\th/(4{\xi}^2)$. To calculate this limit we need to
know the behaviour of $\tilde\th$ near $\xi = 0$ ($\tau = -\ln
2$). The expansion of the system (\ref{ODE1}, \ref{ODE2},
\ref{ODE3}) yields $\tilde\th^{\g + 1} \sim \tilde u -2 \approx  -
2 (1+1/\g) (\tau + \ln 2)$, $r \sim \xi^{-\delta}$, with $\delta =
1/(\g+1)$. This implies that $u(0) = -2/\g$, $\th \sim
\xi^{\delta-2}$. Furthermore, $p \sim \xi^{-2}$, which means that
$P = (x/t)^2 p\, \r_0  \propto g^2 t^2 \r_0$ at $x=0$. The density
diverges as $x \ra 0$ but is integrable (as is necessary to have a
finite mass). Since the self-similar solution is not valid if $x <
c_0^2/g$, we can interpret this divergence as resulting from great
concentration of gas in the region $x \sim c_0^2/g$, relative to
its initial value. In fact, it is easily derived that this mass
ratio grows as $(gt/c_0)^{2\delta}$.

\subsection{Variable gravity}
\label{varg}

If we consider heights of the order of the planet radius or
larger, we have to return to the full continuity equation
(\ref{cont}) and reset $g = GM/x^2$ in the Euler equation
(\ref{Eul}) ($x$ is again the distance from the center). We can
find similarity solutions in the intermediate asymptotics $R \ll x
\ll \R$, in terms of the non-dimensional variable $\xi = x^3/(GM
t^2)$. Hence, we can express the continuity and Euler equations as
\bea
\xi[3u' + (3u-2) {r'\over r}] + 3u = 0,\\
(3u-2)\,\xi u' + u^2 - u = -\xi^{-1} -r^{-1}(2p + 3\xi p'), \label{DE}
\eea
The energy equation becomes
\be
(3u-2)\,\xi [\ln(p/r^\g)]' + 2(u-1) = 0.
\ee

These three equations look similar to the ones corresponding to
constant gravity, and they can also be reduced to two differential
equations for $u(\xi)$ and $\theta(\xi)$, respectively (Cheng
1977, Kazhdan \& Murzina 1994). Unfortunately, they are more
difficult to analyze: the change of variables that transformed the
constant-gravity equations into an autonomous system is not
available. In fact, the free-fall (pressureless) problem is now
nontrivial, although it can be solved.

If $p=0$, Eq.~(\ref{DE}) decouples, becoming an equation for just
$u(\xi)$, namely,
\be (3u-2)\,\xi u' + u^2 - u = -\xi^{-1}. \label{DE-p} \ee
To implement the PDE initial condition, at $\xi \ra \infty$, it is
convenient to make the change of variable ${\hat\xi} = \xi^{-1}$,
transforming the equation into
\be {du \over d{\hat\xi}} = {u^2 - u +{\hat\xi} \over
(3u-2)\,{\hat\xi}}. \label{eq} \ee
The point (0,0) is a singular point of the ODE: a solution is
unspecified, unless we introduce an additional condition. The
initial condition $v=0$ implies that $v=\sqrt{GM/x}
\,{\hat\xi}^{-1/2}u$ goes to zero as ${\hat\xi} \ra 0$. To impose
it, it is best to linearize and solve the ODE around (0,0),
obtaining
\be u \approx C\,\sqrt{{\hat\xi}}-{\hat\xi} ,\label{lin-sol} \ee
so that the singular point is a node. Since $\lim_{{\hat\xi}\ra
0}{\hat\xi}^{-1/2}u = C$, the particular solution of the nonlinear
equation (\ref{eq}) in which we are interested is the only one
that satisfies $u'(0) = -1$ ($C=0$).

\begin{figure}
  \epsfxsize=8cm
\epsfbox{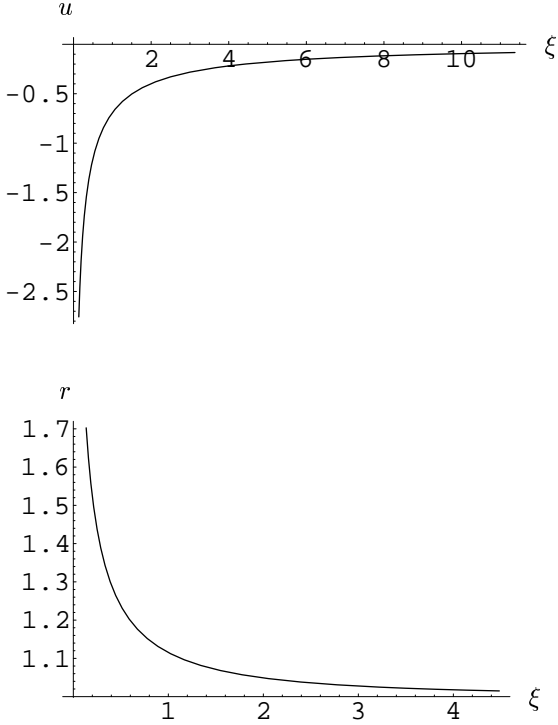} 
\caption{Similarity solution $\{u(\xi),r(\xi)\}$ for free fall
with variable gravity.} 
\label{f-fall}
\end{figure}

The full free-fall solution can be obtained in parametric form
with the Lagrangian formalism (see Appendix 1). It reads
\bea
\xi = {8\cos^{6}\alpha \over  [2\alpha +\sin(2\alpha)]^2}, \label{ffx}\\
u= - (\alpha + {\sin(2\alpha)\over 2})\,\frac{\sin\alpha}{\cos^{3}\alpha},
\label{ffu}\\
r = \frac{8\,\cos^{-3}\a}
  {9\,\cos\a - \cos (3\,\a) + 12\,\a\,\sin \a}, \label{ffr}
\eea
where $\alpha \in (0,\pi/2)$. The corresponding graphs for
$u(\xi)$ and $r(\xi)$ are shown in Fig.~\ref{f-fall}.

This free-fall solution is to be distinguished from the simpler
free-fall solution considered by Cheng (1977) or Kazhdan \&
Murzina (1994), namely, $u = - \sqrt{2^/\xi}$, which corresponds
to vanishing initial energy or, in other words, to the initial
configuration consisting of gas particles at rest at infinite
distance. Note that the dominant term of the solution
(\ref{lin-sol}) is proportional to $\sqrt{1/\xi}$ and coincides
with their free-fall solution for $C = -\sqrt{2}$, being then an
{\em exact} solution of the nonlinear equation (\ref{eq}).
Furthermore, for other non-null values of $C$, we have different
self-similar asymptotic solutions, with a given initial velocity
proportional to $\sqrt{GM/x}$. On the other hand, we can derive $u
= - \sqrt{2^/\xi}$ from the parametric solution in the limit $\xi
\ra 0$ (corresponding to $\a \ra \pi/2$ and meaning small distance
or long time), since the initial energy of the gas particles
becomes negligible in this limit.

The complete similarity solution consists of the free-fall
solution and an inner solution with pressure matching across a
spherical shock wave. The PDE's boundary conditions at $x=0$
determine the form of the inner solution of the two differential
equations for $u(\xi)$ and $\theta(\xi)$ near $\xi = 0$. It must
be a solution of ``type three", with one free parameter (Kazhdan
\& Murzina 1994):
\be u \approx u_0 \xi^{\beta}, \quad \theta \approx
\xi^{-1}\left(1+\theta_0 \,\xi^{\beta}\right), \label{asyKM} \ee
where $u_0$ and $\theta_0$ are related by $$[3(\g-1)\beta + 3\g
-4] u_0 = 2\beta\th_0$$ and $\beta = -1/3 +
(2/3)\sqrt{(\g-1)/\g}$. The match with the previous free-fall
solution across the shock determines the free parameter. There is
no straightforward way to find the shock coordinate $\xi_s$ and a
sort of ``trial and error" procedure is necessary (Chevalier 1989,
Kazhdan \& Murzina 1994). An approximate value can be obtained by
matching the expressions (\ref{asyKM}), namely, $\xi_s = 0.013$.

Both the pre-shock and the post-shock gas velocities are given by
$v(x_s,t) = (GM\xi_s)^{1/3} u(\xi_s)\,t^{-1/3}$, for the
respective $u$ values. Hence, the similarity mass flow across the
shock is ${\dot m} = 4\pi x^2 \r v = 4\pi GM\r_0 \xi_s u(\xi_s)
r(\xi_s)\, t$, which increases with time. This may seem
paradoxical, since the velocity decreases as $t^{-1/3}$, but the
increase in shell mass due to the spherical geometry compensates
for it.

We note that the shock velocity experiences at $x_s \sim R$ a
crossover from $v_s = 2 x_s/t = 2 \xi_s g t$ to $v_s =
2/3\,(x_s/t) = 2/3\,(GM\xi_s)^{1/3} \,t^{-1/3}$ (with {\em
different} $\xi_s$), turning from accelerating to decelerating.
The previous solution is valid for $x_s-R \ll R$, whereas the
present solution is valid for $x_s \gg R$. Once the second
solution takes hold, given that the sound velocity for the
far-away gas is $c_0$, the shock disappears after its velocity
becomes subsonic.  This occurs for $(GM\xi_s)^{1/3} \,t^{-1/3}
\sim c_0 \Rightarrow t \sim GM/c_0^3$ and $x_s \sim \R$.

\section{The static state}

Since the shock wave eventually vanishes, leaving the gas with
more entropy than in the initial state, we expect that it must
reach a stationary state in the long run. The condition that $v(x)
= 0$ (from the continuity equation and the boundary condition)
implies that the stationary solution is actually static.

The hydrostatic equation,
$$\int {dP\over \r} = -\int g \,dx,$$
with $P \propto \r^\g$ and $g$ constant, has the solution:
\be \left({\r\over \r_b}\right)^{\g-1} = 1 - {\g-1\over \g}{\r_b
g\over P_b} (x-R), \label{lawlin} \ee
where the subscript $b$ denotes values at the ball's surface.
Actually, this formula describes well the lower region of the
present Earth's atmosphere. Nevertheless, it is unphysical when
taken over the entire range of $x$ (for $\g > 1$, that is, for a
non-isothermal distribution): $\r^{\g-1}$ (which is proportional
to the temperature) decreases linearly with height, becoming
negative for some height. It is natural that there cannot be a
stationary state with constant gravity, given that nothing can
prevent the free fall of gas.

In contrast, if we take variable $g = GM/x^2$,
\be
\left({\r\over \r_b}\right)^{\g-1} = 1 - {\g-1\over \g}{\r_b g_b R\over P_b}
\left(1-{R\over x}\right).
\ee
In this case, it is more convenient to take the reference at
infinity, where the density and pressure keep their initial values, so that
\be {T\over T_0} = \left({\r\over \r_0}\right)^{\g-1} = 1 +
{\g-1\over \g}{\r_0 \over P_0}{G M\over x} = 1 + (\g-1){\R\over
x}. \label{law} \ee
As expected, $\R$ gives the sphere out of which the gas can be
considered gravitationally unbound (for its thermal plus
gravitational energy is positive) and, consistently, the scale of
crossover to the asymptotic ($x \ra \infty$) density $\r_0$.

For small radius $x \ll \R$, we can neglect the unity in the
right-hand side of Eq.\ (\ref{law}), so that we have power laws.
In particular, the density can be written as
\be
{\r\over \r_0} = \left({\R\over n x}\right)^{n},
\label{power-law}
\ee
by introducing $n$ according to $\g = 1+1/n$. It coincides with
Cheng's hydrostatic solution, after substituting $\r_0 \R^n =
(GM)^n \g^{-n} \r_0^{n+1} P_0^{-n}$. However, Cheng's 0 subscript
indicates {\em arbitrary} reference values, given that $\r^{n+1}
P^{-n}$ is constant. Therefore, we see that it is convenient to
take the asymptotic values as reference, as long as they are non
vanishing, $\R$ being the scale where $\r$ and $P$ take those
values. Moreover, we can let those asymptotic values vanish, by
taking $\R \ra \infty$ while the product $\r_0 \R^n$ has finite
limit. On the other hand, since the ratio between the surface
values of temperature or density and the initial values is given
by $T_b/T_0 = (\r_b/\r_0)^{1/n} = \R/(nR) \gg 1$, we may prefer to
choose the former values as reference. Then we can analogously
take the limit $R \ra 0$ and $\r_b \ra \infty$, while keeping
$\r_b R^n$ constant. Since we have pure power laws when $\R/R \ra
\infty$, the reference is arbitrary.

We must remark that the law (\ref{law}) gives a density contrast
$\r -\r_0$ that decreases too slowly with the radius, yielding a
divergent accumulated mass of gas as $x \ra \infty$.  In fact, the
{\em exact} form of the hydrostatic equation involves the gravity
of the gas and, hence, is equivalent to the equation that
describes stellar structure, namely, the Lane-Emden equation
(Chandrasekhar 1967). If the bound mass of gas is much smaller
than the mass of solid material, that is, if $\R \ll
(M/\r_0)^{1/3}$, the solution of this equation is well
approximated by Eq.~(\ref{law}), while for $x > (M/\r_0)^{1/3}$
the decay is faster than the decay $\sim 1/x$ given by it.  One
can then approximate the mass of accreted gas by the total mass of
gas inside the sphere with radius $\R$, which is larger than the
initially bound gas mass, $\M$, because a portion of the gas that
is initially out of the sphere falls inside.  The detailed
calculation in Appendix 2 shows that the ratio of this portion to
$\M$ is of the order of unity for $\g \leq 4/3$ (in particular, it
takes the value 2.65 for $\g = 7/5$).

We have seen in the previous section that the spherical shock vanishes
when it reaches a radius $\sim \R$. This is consistent with the fact
that the gas outside a sphere of radius $\R$ is inmune to the
gravitational influence of the planet. On the other hand, since the
shock vanishes for $t \sim G M/c_0^3$, this implies that the
similarity solution becomes useless for $t > G M/c_0^3$, marking the
crossover to the static state. Nevertheless, it is necessary to remark
that the static state is an intermediate asymptotic regime not only in
space, as was commented above, but also in time, being always required
that $t \ll 1/\sqrt{G \r_0}$.

\section{Discussion}

We have performed a thorough analysis of a simple model for gas
accretion onto a solid body. We have seen that there are two
relevant length scales, namely, $\R$ and $(M/\r_0)^{1/3}$, the relative
value of which determines whether the dynamics consists of
accretion or collapse: to neglect self gravity, it is required
that $\R \ll (M/\r_0)^{1/3}$.

We have not restricted ourselves to similarity solutions. We have
divided the space dimension (the radius $x$) into a near zone
(near the ball's surface, where $x-R \ll R$ and $g$ is constant)
and a far zone ($x\gg R$). In the near zone, we have obtained an
exact solution for short time and a self-similar
intermediate-asymptotic regime for longer time and radius. In the
far zone, we have a self-similar intermediate-asymptotic regime
for relatively long time and radius and an exact static state for
very long times. These solutions provide a good intuitive
understanding of the entire dynamics.

The crucial feature is the formation and propagation of a shock
wave until its dissipation. The shock wave arises in a short time
$\sim c_0/g$ and it rapidly grows. As long as its propagation is
strongly supersonic, we can consider the gas cold ($c_0 = 0$) or,
in other words, the dynamics driven by gravity. The shock wave
experiences a crossover between one stage of strong dissipation
and acceleration to another of slowing down, owing to decreasing
gravitational energy input, until its eventual vanishing, leaving
behind dense and hot accreted gas in a quasi-static distribution.

The process of accretion begins at $t=0$ and continues until the
end, when $v \ra 0$, but it can be considered finished when $t
\sim GM/c_0^3$. Since the velocity experiences a strong reduction
at the shock, with a consequent increase in the density, it is
sensible to define the accretion rate as the value of the mass
flow ${\dot m}$ there. We have seen that this value is
proportional to $t$ during the second self-similar regime, so the
amount of mass being accreted only begins to decrease at the end
of it. The total accreted mass in the static state is not very
large if $\g \geq 4/3$ ($\g = 4/3$ for the black-body opacity
limit), in spite that the corresponding ratio of surface density
to $\r_0$ is very large ($[\R/(nR)]^n$).

Although an adequate polytropic index accounts for some form of
heat conduction or radiation, the polytropic static state for a
heat conducting or radiating gas is unstable against further
cooling. Therefore, we must remark that the true static state,
from a strictly theoretical thermodynamic point of view, is only
achieved by cooling until reaching an isothermal state. The
corresponding isothermal density distribution is much more
concentrated near the ball's surface (it is exponential), leading
to a far larger accreted mass. However, the time to reach this
isothermal distribution would be, arguably, larger than the gas
collapse time.

We may wonder how our picture would be modified by different
initial conditions, namely, by a non-homogeneous density
distribution (e.g., a power law). Gravity only has effect inside
the sphere of radius $\R$, in which the gas rapidly becomes
supersonic and approaches free fall, until the shock wave arrives.
The free-fall solution (\ref{ffx},\ref{ffu},\ref{ffr}) is
independent of the value of $\r_0$, which can be a function of $x$
(see Appendix 1), but the form of the shock wave and the
post-shock gas distribution will change.

We may also consider an initial non-homogeneous temperature
distribution. Near the body surface, we can make a linear
approximation of it. If its slope is sufficiently smaller (in
absolute value) than the static value provided by Eq.\
(\ref{lawlin}), we can still consider it isothermal and predict
that the fall of gas will produce a shock wave. Nevertheless, we
can envisage distributions close to the static one such that the
gas falls gently on the surface without ever giving rise to a
shock wave (such as, e.g., the distribution corresponding to the final
stage of our process, after the vanishing of the shock wave).
Presumably, these distributions lead to little accretion.

Now, I briefly estimate the numerical scales suitable for the
application to neutron star accretion or planet accretion. For a
neutron star of 1 $M_\odot$ with $R \simeq 10^4$ m, $g \simeq
10^{12}$ m s$^{-2}$. The sound speed (proportional to the square
root of the temperature) can be large, say $c_0 \simeq$ 10000 m
s$^{-1}$. Then, $c_0/g \simeq 10^{-8}$ s and $c_0^2/g \simeq
10^{-4}$ m, $R^{3/2}/\sqrt{GM} = \sqrt{R/g} \simeq 10^{-4}$ s, but
$\R \simeq 10^{12}$ m and $GM/c_0^3 \simeq 10^8$ s. It is not
clear how to determine $\r_0$. A ``typical" galactic value would
give a collapse time similar to star formation times, that is, of
the order of a million years. Considering a planet placed in the
protoplanetary nebula, we may take for $g$ the Earth value of 10 m
s$^{-2}$ and, for example, $c_0 \simeq$ 300 m s$^{-1}$ (its
current value in the low Earth's atmosphere); the initial typical
scales of the problem, namely, the time $c_0/g$ and the distance
$c_0^2/g$, take values of 30 s and 9 Km, respectively. The basic
time scale is $\sqrt{R/g} \simeq 800$ s (for an earth-like planet
with $R \simeq 6000$ Km).  We further estimate $\R \simeq GM/c_0^2
\simeq 5~10^6$ Km and $GM/c_0^3 \simeq 5~10^7$ s $\simeq 1$ yr.  A
plausible value for the density of the protoplanetary nebula is
$\r_0 \sim 10^{-11}$ g/cm$^3$. We see that its collapse time
$1/\sqrt{G \r_0} \simeq 10^9$ s is sufficiently larger than
$GM/c_0^3$ for neglecting self gravity.

\subsection*{Acknowlegments}
I thank A.~Mancho, J.M. Mart{\'\i}n-Garc{\'\i}a,
J.~P\'erez-Mercader and M.P. Zorzano for conversations,
A.~Dom{\'\i}nguez for comments on the manuscript, and
R.~G\'omez-Blanco for reading a preliminary version of the
manuscript. This work is supported by a ``Ram'on y Cajal'' 
contract and by grant BFM2002-01014 of the Ministerio de Ciencia y
Tecnolog\'{\i}a.

\appendix

\subsection*{Appendix 1: Lagrangian similarity solution for the free fall in the field of
a point-like mass}

We describe here the ``Lagrangian solution'' of Eq.~(\ref{DE-p}),
\be (3u-2)\,\xi u' + u^2 - u = -\xi^{-1}. \label{eq1}
\ee
In the Lagrangian picture, the velocity is given by the energy
equation
\be {v^2\over 2} = E + {GM\over x} \ee
(because the energy is conserved), that is,
\be v = -\sqrt{2(E+ {GM\over x})}. \label{v} \ee
The initial condition that the particle is at rest implies that $E
= -GM/x_0$. Hence,
\be u = {t\over x}v = -\sqrt{2\,{\xi}^{-1}(1- {x\over x_0})},
\label{u} \ee
and to have a self-similar form we must express $x/x_0$ as a
function of $\xi$. In order to do it, we must solve the equation
of motion (\ref{v}):
\be t = -\int_{x_0}^x {ds\over \sqrt{2({GM/s}-{GM/x_0})}}. \ee
First, we write it in a self-similar form, by introducing $\s =
s/x_0$,
\be t = -{x_0^{3/2}\over \sqrt{GM}} \int_1^{x/x_0} {d\s\over
\sqrt{2({1/\s}-1)}}. \label{int} \ee
It is clear now that
\be \xi^{-1/2} = F(x/x_0), \label{F} \ee
where
$$F(y) = -y^{-3/2}
\int_1^{y} {d\s\over \sqrt{2({1/\s}-1)}}.$$ Inverting (\ref{F}) we
obtain $x/x_0$ as a function of $\xi$, which upon substitution in
(\ref{u}) yields its self-similar form. This is the solution of
the ODE that satisfies the appropriate initial condition (see
section \ref{varg}).

The method exposed above works in general for free self-similar
motion in an external gravitational field. In the present case,
one can calculate the integral in Eq.~(\ref{int}) by the change of
variables $\s = \cos^2 \varphi$. This allows one to obtain the
solution of the differential equation (\ref{eq1}) in parametric
form: Let
\be \cos^2 \alpha = x/x_0; \label{defa} \ee
then,
\be {\sqrt{GM}\, t\over {x_0}^{3/2}} = {1\over \sqrt{2}}\,(\alpha
+ {\sin(2\alpha)\over 2}),  \label{ta} \ee
and
\bea \xi^{-1/2} = {1\over \sqrt{2}} \cos^{-3}(\alpha) \,(\alpha +
{\sin(2\alpha)\over 2}),  \label{xia}\\
u= -\cos^{-3}(\alpha) \,(\alpha + {\sin(2\alpha)\over 2})\,\sin\alpha,
\eea
where $\alpha \in (0,\pi/2)$. It is easy to verify that this
parametric solution fulfills the initial condition
$\lim_{\xi\ra\infty}u' = -1$.

In the Lagrangian formulation, the density is given by (Landau \&
Lifshitz 1987) $$\r = \left(\partial x_0\over \partial
x\right)_{\! t} \frac{x_0^2}{x^2}\, \r_0,$$ where $x_0^2/x^2$ is a
spherical geometry factor. From Eqs.~(\ref{defa}), (\ref{ta}) and
(\ref{xia}),
\be r= {\r\over \r_0} = \frac{8\,\cos^{-3}(\a)}
  {9\,\cos (\a) - \cos (3\,\a) + 12\,\a\,\sin (\a)},
\ee
which together with Eq.~(\ref{xia}) constitute the parametric
equations of the density.

\subsection*{Appendix 2: Accreted mass in the static solution}

Let us calculate the gas mass enclosed in the sphere of radius
$\R$ in the final static state:
\bea
\M_f = \int_R^\R \r(x)\, 4\pi x^2 dx =\nonumber\\
4\pi\r_0 \int_R^\R \left(1 + (\g-1){\R\over x}\right)^{1\over\g-1}
x^2 dx
\eea
This integral is convergent for $R \ra 0$ if $\g > 4/3$. We can
easily express it in non-dimensional form:
\be
\M_f = 4\pi\r_0 \R^3 \int_{R/\R}^1 \left(1 + (\g-1){1\over
s}\right)^{1\over\g-1} s^2 ds.
\ee
Now, it is convenient to make the change $\g = 1+1/n$. The
resulting integral can be computed in terms of a hypergeometric
function:
\bea
\int_{R/\R}^1 \left(1 + {1\over n s}\right)^{n} s^2 ds
=\nonumber\\
\left[{{\frac{s^{3-n}}{(3-n) n^n}}}\,
{}_2F_1(3 - n,-n,4 - n;-ns) \right]_{R/\R}^1.
\eea

\begin{figure}
  \epsfxsize=8cm
\epsfbox{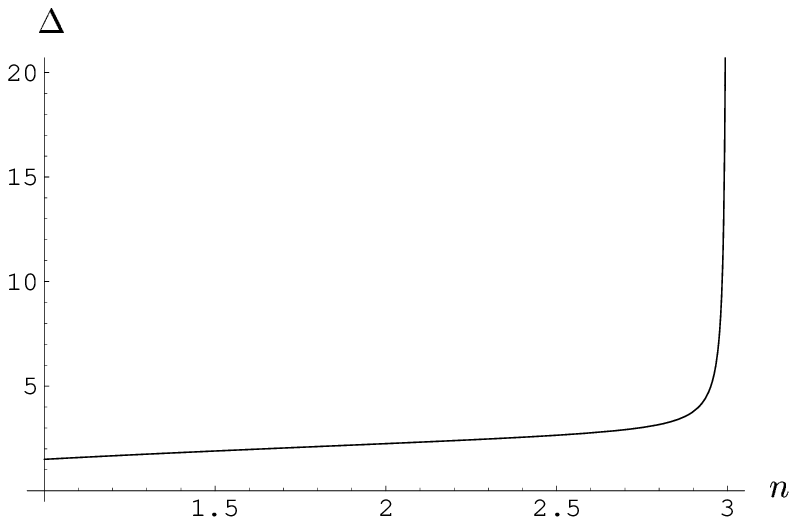} 
\caption{Accreted mass ratio $\Delta$ as a function of $n$ ($\g =
1+1/n$).} 
\label{mass}
\end{figure}

We are interested in the limit of $[\M_f-\M]/\M$ when $R/\R \ra
0$, in which we are left with the value of the bracket at $s=1$.
Denoting $\Delta = \lim_{R/\R \ra 0} \M_f/\M -1$,
\bea
\Delta = 3 \int_0^1 \left(1 + {1\over ns}\right)^{n} s^2 ds
-1
=\nonumber\\
={{\frac{3}{(3-n) n^n}}}\,
{}_2F_1(3 - n,-n,4 - n;- n) -1.
\eea
The value of $\Delta$ grows almost linearly, with a small slope,
up to near the pole $n=3$ ($\g = 4/3$), as Fig.~\ref{mass} shows.
In particular, it takes the value 2.65099 for $\g = 7/5$
($n=2.5$).

The pole shows the divergence of the integral as $s \ra 0$. Then,
it is not possible to make $R/\R = 0$ near the pole. The $n=3$
integral yields
\be
\Delta = 3 \int_{R/\R}^1 \left(1 + {1\over 3s}\right)^{3} s^2
ds -1 =
\frac{1}{9} \ln{\R \over R} + \frac{5}{2}\,. \ee
Its value is smaller than 5 even for $\R/R$ as large as $10^{9}$.

\end{document}